# On the representation of dense plasma focus as a circuit element

S K H Auluck[1]

International Scientific Committee for Dense Magnetized Plasmas,
http://www.icdmp.pl/isc-dmp
Hery 23, P.O. Box 49, 00-908 Warsaw, Poland

Abstract: The dense plasma focus is a plasma discharge powered by a capacitor bank. Standard diagnostics include measurement of the time derivative of the current through and the voltage across its connections with the capacitor bank. Interpretation of this diagnostic data often involves some assumptions regarding the representation of the dense plasma focus as a time varying inductance. One of the characteristic features of the current derivative waveform is a relatively sharp dip and an associated sharp voltage spike. This has often been interpreted as a result of a rapid rise in the time varying inductance of the plasma. Sometimes, an anomalous plasma impedance is invoked. This Letter discusses instances where such interpretation creates conceptual difficulties. A first principles approach to the representation of the dense plasma focus as a circuit element reveals some fundamental problems with the traditional representation of plasma focus as a time varying inductance. The anomalous impedance is shown to be necessary to account for the difference in the motional impedance implied by a time-varying inductance in the circuit element representation and a first principles description based on Poynting's Theorem. Dynamo effects that convert post-stagnation local motion of plasma into 3-dimensional magnetic fields are shown to contribute to the effective inductance of the plasma focus and resolve the observed conceptual difficulties.

The Dense Plasma Focus [1,2] is well-known as an inexpensive intense source of neutrons, fast ions, electrons and x-rays[2-6], a rich repository of exotic plasma phenomenology [2], a novel environment for synthesis of new materials [7] and a versatile educational tool for training young plasma physicists[8]. But at its most basic level, it is a plasma discharge powered by a capacitor bank. Its most common diagnostic is measurement of the time derivative of current through its terminals, closely followed by measurement of voltage across its connections with the capacitor bank. Occurrence of a sharp dip in the current derivative signal and a corresponding sharp spike in the voltage signal are taken as indications of a proper operation of the device. Parameters of these transients are found to be correlated with intensity and yield of neutron emission [9].

---

[1] Corresponding Author. email: skhauluck@gmail.com

Interpretation of this diagnostic data[1,2,9,10] involves representing the plasma discharge as a circuit element, usually as a time varying inductance, in a circuit schematic that can be found in the cited references and is therefore not duplicated here. Evaluation of this "interpreted inductance" for PF-1000 [11,12] shows a quantity monotonically increasing with time, an observation also confirmed on smaller devices [13,14]. Sometimes, the voltage spike, that is time-correlated but not simultaneous [15] with the sharp current derivative minimum, is interpreted as being the result of an anomalous impedance[2] ascribed to micro-instabilities.

The monotonically increasing "interpreted inductance" presents conceptual difficulties in associating a physical mechanism with it which are discussed below.

The usual physical picture involves a current front that slides along the anode surface and converges on the axis in a funnel-like shape causing a rapid increase in inductance. This is supported by experiments on PF-1000 [16,17] which utilize magnetic probes placed through the anode and correlated with 15-frame interferometry. In these experiments, interferometry pictures show contours of equal areal electron density (integration of electron density along the chord of the plasma cross-section in the path of the laser beam) at intervals of ~10-15 ns. Absolutely calibrated magnetic probes placed at radii 40, 13 and 0 mm with the sensitive element 10 mm above the anode surface showed the following noteworthy features:

- The current derivative minimum was taken as the origin of time.

- Almost all the discharge current that entered the device was carried past the 40 mm radius in a layer of thickness 1.6-2.6 cm that lay a few millimeters behind the dense sheath visible in the interferogram. This was based on the estimate of average velocity of the sheath ~$2.1 \times 10^5$ m/sec with a 25% shot-to-shot variation. The boundary between the dense plasma and the current carrying layer in the reported pictures is quite sharp in the sense that the density falls by at least two orders of magnitude within a radial interval less than 1 mm. This is often taken as confirmation of the idea that the dense plasma is driven by a magnetic piston, at least during a major portion of the plasma propagation time. .
- In one illustrative shot, interferograms at -68, -38 and 22 ns show (Fig 6 of Ref 17) the position of the plasma with respect to the probe which correspond to the start, maximum and zero of the probe signal at the 13 mm radial position. The leading edge of current density front at -68 ns lies ahead (towards the axis) of the leading edge of the hollow dense plasma shell. The maximum of the probe signal, that corresponds to the maximum of current density, lies at the outer edge of the dense column at -38 ns. The zero of the probe signal at 22 ns corresponds to the peak of the magnetic field and the interferogram shows an already expanded plasma column. *Most of the current density must then be already located inside the dense expanding plasma at this time*.
- Even taking into account a reasonable value of ~10-20 ns for the propagation delay between occurrence of a current derivative minimum at the front of the anode and its detection ~2 m away outside the large plasma focus installation [1], it is clear that the current derivative minimum and the arrival of current density within the dense plasma column are simultaneous to within experimental error. It is, therefore, a reasonable conjecture that the current derivative minimum is *caused* by the diffusion of the current density to the axis, which leads to a rapid increase in inductance that varies as the logarithm of the radius of the current distribution [2,12], which is expected to cause a corresponding decrease in the current.

- The increase of "interpreted inductance" beyond the current derivative minimum is not so simple to understand. Inductance as a property of a circuit element is usually understood in terms of magnetic flux per unit current or twice the total magnetic energy divided by square of the current. This presumes a geometry of the conductor. After the breakup of the plasma column, a defined geometry of current flow is not discernible. What, then, is the physical meaning of the post-breakup monotonically increasing "interpreted inductance" ?

The conjecture regarding the causal relation between arrival of the current front to the axis and the occurrence of the current derivative minimum also runs into difficulty when compared with other data. On PF-1000, pure neon plasma focus operation[18] shows the current derivative minimum occurring more than 200 ns after the formation and breakup of the dense plasma column as seen by interferometry and confirmed by intense soft x-ray emission. If the dense plasma is really driven by the magnetic piston, the current front should be close behind the dense plasma and should reach the axis shortly after formation of the dense column. The observed delay of more than 200 ns between the plasma column breakup and current derivative minimum suggests three possibilities: either (1) the current sheath is not attached to the dense plasma and is far behind it (suggesting that the dense plasma is *not* driven by the magnetic piston), or (2) it diffuses much more slowly in neon plasma as compared with deuterium plasma (contrary to current understanding of resistive diffusion) or (3) the current derivative minimum is not *caused* by the transport of the current to the axis. The first two possibilities appear less likely than the third because these phenomena are relatively better grounded in physics. The third possibility tries to make a correspondence between a circuit element description of the plasma focus in terms of a total device current and its physical origin in terms of propagation of a current front. Dychenko and Immshennik [19] have carried out a detailed analysis of the MHD foundations of the plasma focus problem, in which they make assumptions that the axial and radial components of magnetic field and azimuthal component of plasma velocity are initially zero. Their analysis concludes that if the azimuthal symmetry is maintained, these components continue to remain zero. This assumption precludes dynamo actions that could convert kinetic energy of plasma into energy of 3-dimensional (3-D) magnetic field. The detection of axial component of magnetic field in PF-1000 [17] and its correlation with neutron emission therefore calls for a re-examination of the third possibility. This Letter takes a closer look at the representation of the plasma focus as a circuit element.

A circuit element is defined as an idealized two terminal object with a certain voltage V(t) across it and a certain current I(t) passing through it. The voltage is defined as the work done in transporting a test charge *against the electric field* between the terminals divided by the test charge. While constructing an idealized circuit element representation of the plasma focus, or in general any plasma in contact with two electrodes labeled 1 and 2, it must be borne in mind that the streamline of the electric field starting from one point on electrode 1 and ending on another point on electrode 2 may make a complicated spatial excursion both inside and outside the plasma. The presence of complex motional dynamo effects that amplify a non-zero seed magnetic field by converting kinetic energy of the plasma into magnetic energy implies that the electric field lines could even turn several times around the axis or make multiple off-axis loops. In general, the streamlines connecting one pair of points on the two electrodes may have a very different spatial structure from another pair of points. Hence while representing the spatially extended electrodes as point terminals of an idealized circuit element, the voltage $V_{12}$ must be considered as the *average* work done while transporting the test charge between a pair of points on the two electrodes along the streamlines of electric field connecting the pair. This provides the following physical basis for constructing a circuit element representation:

$$V_{12}(t) = -\frac{1}{I(t)}\int_\Omega d^3\vec{r}\, \vec{J}\cdot\vec{E} \qquad (1)$$

where the right hand side is the total electric power flowing through the two terminals expressed in terms of fields inside the device divided by the total current flowing through the device. *Every physical phenomenon, without exception,* that happens within the plasma chamber occurs because of the power and current supplied from outside. The expression for the voltage across the terminals in a circuit representation must allow the power required by each of such physical phenomena to be drawn from the electrical circuit in an adequate manner.

This 3-D spatial integration is over a domain $\Omega$ such that $\vec{J}$ is zero outside it. Excluded from this domain is the interface between the 'circuit element' and the external power source, through which, the power enters the device. In the case of the plasma focus, this would be the cathode plate that is in contact with the insulator and the squirrel cage, or its smaller portion in the initial phase.

Physical systems do not have sharply defined boundaries with an arbitrarily fine spatial scale. Interferometry pictures (for example, Fig 6 of Ref 17) show that the dense plasma has a boundary region where the density falls over two orders of magnitude over a radial distance less than 1 mm. Hence the domain of integration in (1) is idealized with a bounding surface that is smeared over a finite sub-millimeter-scale transition region in reality. This bounding surface, designated $\Sigma$, comprises the contact surfaces between the plasma and each of the two electrodes, through which the current passes in, as well as the extreme boundary of the plasma bridging the two electrodes that separates the current carrying region of the plasma from the current free region. The domain $\Omega$ is not a simply-connected domain for a plasma focus before breakup of the plasma column. Rather, topologically it is a toroid. This is illustrated in Fig 1.

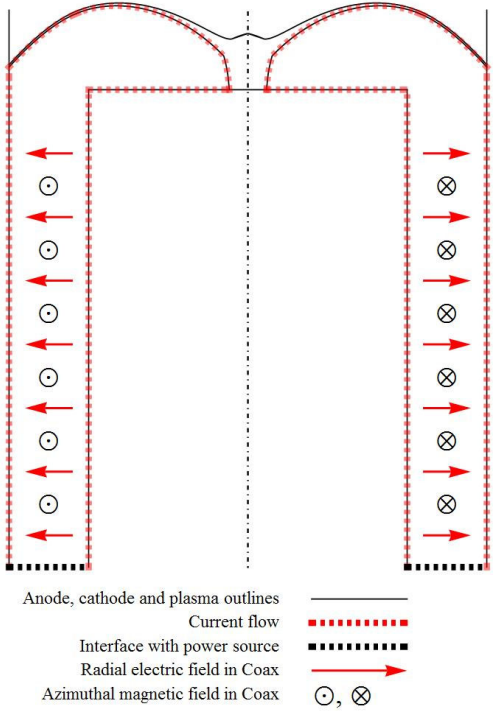

Fig. 1: Domain $\Omega$ is the volume enclosed by the current-carrying surfaces of the electrodes and the plasma depicted by red dashed lines and the interface with the power source depicted by the black dashed lines. It is seen to be a topological toroid: there exist simple curves (those encircling the axis) that cannot be continuously deformed into a point. Electric field $\vec{E}$ in the coaxial region is radially directed, shown by red arrows and the magnetic field $\vec{B}$ is in the azimuthal direction. The Poynting vector $\mu_0^{-1}\vec{E}\times\vec{B}$ is directed along the axial direction. Its surface integration over the black-dashed interface with the power source can be easily shown to be equal to $I(t)\cdot V(t)$

Using Poynting's theorem,

$$\vec{J}\cdot\vec{E} = -\frac{\partial}{\partial t}\left(\tfrac{1}{2}\varepsilon_0 E^2 + \tfrac{1}{2}\mu_0^{-1}B^2\right) - \mu_0^{-1}\vec{\nabla}\cdot\left(\vec{E}\times\vec{B}\right) \qquad (2)$$

the integral in (1) can be written in terms of total time derivative of spatially integrated quantities

$$\begin{aligned}
I(t)V(t) &= \int_\Omega d^3\vec{r}\,\frac{\partial}{\partial t}\left(\tfrac{1}{2}\varepsilon_0 E^2 + \tfrac{1}{2}\mu_0^{-1}B^2\right) + \int_\Omega d^3\vec{r}\,\mu_0^{-1}\vec{\nabla}\cdot\left(\vec{E}\times\vec{B}\right) \\
&= \frac{d}{dt}\int_\Omega d^3\vec{r}\left(\tfrac{1}{2}\varepsilon_0 E^2 + \tfrac{1}{2}\mu_0^{-1}B^2\right) - \int_{\Sigma_P} d\vec{S}\cdot\vec{v}\left(\tfrac{1}{2}\varepsilon_0 E^2 + \tfrac{1}{2}\mu_0^{-1}B^2\right) + \mu_0^{-1}\oint_\Sigma d\vec{S}\cdot\left(\vec{E}\times\vec{B}\right)
\end{aligned} \qquad (3)$$

The second integral is evaluated only on the moving boundary $\Sigma_p$ of the domain $\Omega$ since stationary boundaries do not contribute to it. The third integral is evaluated over the entire surface $\Sigma$. Since the geometry of the integration domains is a function of time, the three integrals in (3) would be evaluated differently for each phase of the plasma focus. After the rundown phase, the electric field in the coaxial region is along the radial direction, the magnetic field is in the azimuthal direction so that the Poynting vector $\mu_0^{-1}\vec{E}\times\vec{B}$ is along the axial direction (see Fig 1). Its surface integral over the coaxial electrodes is therefore zero. Along the cathode plate at the bottom, which is excluded from the domain for being the interface between the circuit element and the power source, the surface integral is exactly equal to the power input $I(t)V(t)$.

Using the Generalized Ohm's Law

$$\vec{E} = -\vec{v}\times\vec{B} + \eta\vec{J} \tag{4}$$

the part of the third integral on $\Sigma_p$ can be written as

$$\mu_0^{-1}\oint_{\Sigma_p} d\vec{S}\cdot(\vec{E}\times\vec{B}) = \mu_0^{-1}\oint_{\Sigma_p} d\vec{S}\cdot\vec{v}(\vec{B}\cdot\vec{B}) - \mu_0^{-1}\oint_{\Sigma_p} d\vec{S}\cdot\vec{B}(\vec{B}\cdot\vec{v}) + \mu_0^{-1}\oint_{\Sigma_p} d\vec{S}\cdot\eta\vec{J}\times\vec{B} \tag{5}$$

Equation (3) then becomes

$$I(t)V(t) = \underbrace{\frac{d}{dt}\int_{\Omega} d^3\vec{r}\left(\tfrac{1}{2}\mu_0^{-1}B^2\right)}_{\text{I}} + \underbrace{\int_{\Sigma_p} d\vec{S}\cdot\vec{v}\left(\tfrac{1}{2}\mu_0^{-1}B^2\right)}_{\text{II}} + \underbrace{\frac{d}{dt}\int_{\Omega} d^3\vec{r}\left(\tfrac{1}{2}\varepsilon_0 E^2\right)}_{\text{III}} - \underbrace{\int_{\Sigma_p} d\vec{S}\cdot\vec{v}\left(\tfrac{1}{2}\varepsilon_0 E^2\right)}_{\text{IV}}$$
$$+ \underbrace{\mu_0^{-1}\oint_{\Sigma_p} d\vec{S}\cdot\eta\vec{J}\times\vec{B}}_{\text{V}} - \underbrace{\mu_0^{-1}\oint_{\Sigma_p} d\vec{S}\cdot\vec{B}(\vec{B}\cdot\vec{v})}_{\text{VI}} \tag{6}$$

One can compare various labeled terms in (6) with the expression for power through the circuit elements defined as inductance, capacitance and resistance, whose geometrical parameters could be time-dependent.

$$P_L = I\frac{d}{dt}(LI) = \frac{d}{dt}\left(\tfrac{1}{2}LI^2\right) + \tfrac{1}{2}I^2\frac{dL}{dt} \; ; \tag{7}$$

$$P_C = V\frac{dQ}{dt} = V\frac{d}{dt}(CV) = \frac{d}{dt}\left(\tfrac{1}{2}CV^2\right) + \tfrac{1}{2}V^2\frac{dC}{dt} \tag{8}$$

$$P_R = I.V_R = RI^2 \tag{9}$$

The terms I and III in (6) are time derivatives of the total magnetic and electric energy that clearly must correspond to $d/dt\left(\tfrac{1}{2}LI^2\right)$ and $d/dt\left(\tfrac{1}{2}CV^2\right)$ respectively.

The terms proportional to $dL/dt$, which is sometimes called as the motional impedance, in (7) and $dC/dt$ in (8) are clearly dependent on the velocity with which the dimensions of the inductance and capacitance are changing and also to the square of the current and voltage respectively. Terms II and IV have this property.

The term V in (6) is proportional to the resistivity and to the square of the current and is independent of the velocity. This property is shared by the expression for power through a resistance as given in (9).

The next course of action would be to define the "plasma inductance" as

$$L_p \equiv \frac{1}{I^2}\int_\Omega d^3\vec{r}\left(\tfrac{1}{2}\mu_0^{-1}B^2\right) \tag{10}$$

the "plasma capacitance" as

$$C_p \equiv \frac{1}{V^2} \int_\Omega d^3\vec{r} \left( \tfrac{1}{2} \varepsilon_0 E^2 \right) \tag{11}$$

and "plasma resistance" as

$$R_p = I^{-2} \mu_0^{-1} \oint_{\Sigma_p} d\vec{S} \cdot \eta \vec{J} \times \vec{B} \tag{12}$$

But this would run into a problem because in general,

$$\tfrac{1}{2} I^2 \frac{dL_p}{dt} \neq \int_{\Sigma_p} d\vec{S} \cdot \vec{v} \left( \tfrac{1}{2} \mu_0^{-1} B^2 \right) \tag{13}$$

$$\tfrac{1}{2} V^2 \frac{dC_p}{dt} \neq -\int_{\Sigma_p} d\vec{S} \cdot \vec{v} \left( \tfrac{1}{2} \varepsilon_0 E^2 \right) \tag{14}$$

The difference between the term $\tfrac{1}{2} I^2 \, dL_p/dt$ and term II in (6) would have to be accounted for by invoking an "anomalous impedance" – an impedance that has no analog in circuit theory. The "plasma capacitance" and related term IV in (6) would probably be numerically negligible in a circuit equation in comparison with the capacitance of the energy storage but could nonetheless have detectable effects such as oscillations of current and induced charges on nearby floating conductors.

Term VI depends on the component of velocity along the magnetic field and on the magnetic field component normal to the plasma surface. In a conventional description of the plasma focus, where the magnetic field is purely azimuthal and the azimuthal component of velocity is zero, this term would be zero. But when the PF-1000 phenomenology involving axial magnetic field in the radial phase is considered along with the clearly present axial motion of the plasma, this term becomes non-zero. During stagnation of the radially collapsing plasma near the axis, the radial momentum and associated kinetic energy must be conserved in spite of the fact that radial motion of the plasma cannot proceed further. In an ideally symmetric, purely hydrodynamic case, this would cause a pressure spike on the axis and a radially diverging shock wave. For a magnetized plasma, conservation of radial momentum and associated kinetic energy can happen by creation of turbulent eddies having a 3-dimensional velocity distribution which would also drive a motional dynamo amplifying initial seed magnetic field into a 3-D magnetic field structure. This would ensure that $(\vec{B} \cdot \vec{v})$ is not identically zero. Because of a poloidal magnetic field present in the umbrella-like plasma shape which forms part of the surface $\Sigma_p$, it cannot be assumed that the magnetic field is everywhere parallel to $\Sigma_p$ which would have made the integrand in term VI identically zero. Term VI apparently has no analog in circuit theory and would thus have to be also accounted for in terms of an anomalous impedance.

Because of dynamo activities, current streamlines may condense into local quasi-closed formations like balls of wool with free ends, that can transport the current from the anode to the cathode even after plasma breakup. Such "balls of current streamlines" would be visible as plasma structures in VUV[18] and interferometry images[17] and be manifested as ball-like fast ion traps in reaction product images [20,21].

The monotonic increase of "interpreted inductance" [11,12,13,14] after the pinch breakup and the delay between the pinch breakup and current derivative minimum in a neon plasma focus [18] both become understandable in terms of the above discussion. Notice that the plasma inductance (10) is not limited to the magnetic field generated by the current that flows in the external circuit but involves all three components of the magnetic field. This includes the poloidal magnetic field generated by a motional dynamo as well as any smaller scale eddies, which would arise as a result of conservation of the radial momentum density and associated kinetic energy. In terms of a circuit element description, this would correspond to an increase in the toroidal component of current indicating that the number of turns of the current path around the axis is increasing. The delay between the plasma column breakup and current derivative minimum in a neon plasma [18] can be explained in terms of higher inertia of neon leading to a slower dynamo that takes a longer time for amplifying the seed magnetic field into a poloidal magnetic field component comparable in magnitude to the toroidal magnetic field component, which would cause a rapid rise in inductance and corresponding decrease in current. The occurrence of a hard x-ray peak simultaneous with the current derivative minimum in the neon plasma focus[18] is evidence of significant magnetic activity occurring at the time of the current derivative minimum.

Indications of such post-pinch MHD activity resulting in generation of local plasma structures ~~was~~ were reported by Neil and Post[22] using $CO_2$ laser scattering. A simple experimental test that detects azimuthal currents generated by dynamo activity was recently proposed [23] and demonstrated [24].

In conclusion, this Letter looks at the representation of the plasma focus as a circuit element from a field point of view in terms of Poynting's Theorem. The attempt to define the plasma inductance in terms of total magnetic energy is shown to create an incomplete description of the motional impedance related to the time varying inductance. The difference between the actual term from Poynting's theorem and the motional impedance calculated from the plasma inductance defined from total magnetic energy must then appear as an "unaccounted" impedance.

An additional term, that apparently has no analog in circuit theory, appears in the presence of a 3-D magnetic field structure accompanied with a 3-D velocity structure. The monotonic increase of interpreted inductance after the plasma column breakup and delay between the plasma column breakup and the current derivative singularity in a neon plasma focus becomes comprehensible in terms of increase in the toroidal component of current density as a result of motional dynamo.

The conjecture in Ref 19, p.222, about possible existence of toroidal current loops that have "..no electromagnetic coupling with the external circuit" is physically incorrect since such current loops need energy to evolve and decay that can enter the device only through its connections with the external circuit. The error lies in the fact that the current loop is not an independent entity as presumed in Ref 19 but merely a topological property of the current that enters and exits the device through its terminals.

The author would like to acknowledge Dr. Ryszard Miklaszewski for significant help.

Data Availability Statement: " Data sharing not applicable – no new data generated"

24. Mladen Mitov, Alexander Blagoev, Stanislav Zapryanov and S K H Auluck, "First dedicated experiment on the azimuthal electric field as a new diagnostic tool for Dense Plasma Focus", 9 th International Workshop and Summer School on Plasma Physics, Sofia, Bulgaria, 30th November- 4th December 2020.